\documentclass[aps,prl,twocolumn]{revtex4-2}%
\usepackage[]{graphicx}
\usepackage{xcolor}%
\newcommand{\chem}[1]{\ensuremath{\mathrm{#1}}}

\begin{document}
\preprint{APS/123-QED}

\title{Efficient adiabatic demagnetization refrigeration to below 50 mK with UHV compatible Ytterbium diphosphates $A$YbP$_2$O$_7$ ($A=$Na, K)}

\author{U. Arjun$^{1,2}$}
\email{arjunu@iisc.ac.in}

\author{K. M. Ranjith$^3$}

\author{A. Jesche$^2$}
\email{anton.jesche@physik.uni-augsburg.de}

\author{F. Hirschberger$^2$}
\author{D. D. Sarma$^1$}
\author{P. Gegenwart$^2$}
\email{philipp.gegenwart@physik.uni-augsburg.de}

\affiliation{$^1$Solid State and Structural Chemistry Unit, Indian Institute of Science, Bangalore-560012, India}
\affiliation{$^2$Experimental Physics VI, Center for Electronic Correlations and Magnetism, Institute of Physics, University of Augsburg, 86135 Augsburg, Germany}
\affiliation{$^3$Laboratoire National des Champs Magn\'{e}tiques Intenses-EMFL, CNRS, Université Grenoble Alpes, 38042 Grenoble, France}

\date{\today}

\begin{abstract}
Attaining milli-Kelvin temperatures is often a prerequisite for the study of novel quantum phenomena and the operation of quantum devices. Adiabatic demagnetization refrigeration (ADR) is an effective, easy and sustainable alternative to evaporation or dilution cooling with the rare and super-expensive $^3$He. Paramagnetic salts, traditionally used for mK-ADR, suffer from chemical instability related to water of crystallization. We report synthesis, characterization as well as low-temperature magnetization and specific heat measurements of two new UHV compatible candidate materials NaYbP$_2$O$_7$ and KYbP$_2$O$_7$. Utilizing the PPMS at 2~K, the ADR of sintered pellets with Ag powder admixture starting at 5~T yields base temperatures (warm-up times) of 45 mK (55 min) and 37~mK (35 min) for NaYbP$_2$O$_7$ and KYbP$_2$O$_7$, respectively, slightly advantageous to  KBaYb(BO$_3$)$_2$ (45 mK and 40 min) studied under similar conditions.
\end{abstract}
\maketitle

\section{Introduction}

Quantum effects are most evident at low-temperatures where the thermal fluctuations are suppressed. Novel quantum effects such as Bose-Einstein condensation~\cite{zapf563}, superconductivity~\cite{onnes113}, quantum Hall effect~\cite{Goerbig1193}, quantum spin liquid~\cite{Zhou025003}, quantum spin-ice ~\cite{bramwell374010}, etc. have been experimentally discovered only at temperatures close to absolute-zero. The global scarcity of helium~\cite{kelley8,olafsdottir1} and the increasing need for refrigeration for various technological applications enhanced the importance of helium-free magnetic refrigeration techniques. The isotope $^{3}$He, used in these refrigeration techniques, is extremely scarce in availability and thus extraordinarily expensive due to its increased demand in the defence sector for the production of neutron detectors to combat nuclear terrorism~\cite{kouzes2010,shea2011,Cho778}.

The significant temperature changes in some magnetic materials due to the magneto-caloric effect (MCE) while exposed to an external field~\cite{weiss103} make them useful for attaining low temperatures through adiabatic demagnetization refrigeration (ADR)~\cite{debye1154, giauque768,pobell2007}. Currently milli-Kelvin (mK) ADR is intensively used in satellites and other space technologies~\cite{shirron130,shirron581,shirron915,luchier591,hagmann303} and has already been shown to be potentially practical for quantum computers~\cite{jahromi2019}. More generally, MCE-based magnetic refrigeration significantly reduces environmental impact and offers energy savings of nearly 30\% compared to conventional techniques that use refrigerant gases~\cite{Alahmer4662}.

For ADR applications, replacing a $^{3}$He refrigerator or a $^{3}$He-$^{4}$He dilution refrigerator, materials with a significant entropy change $\Delta S$ in the temperature range between 10 mK and 4 K are required. They should also exhibit a strong MCE within the lowest possible applied magnetic field. In the multi-stage process for attaining sub-Kelvin temperatures, the material is first pre-cooled to a temperature of about $2$~K in a magnetic field of typically a few teslas (e.g. using a pumped $^{4}$He bath or a pulse-tube cooler). Subsequently, the thermal contact with the bath is cut by pumping out the He-gas or opening a heat switch, and the cooling substance is demagnetized. In this adiabatic process, the very low entropy of the cooling substance due to pre-cooling in the magnetic field requires that the temperature of the cooling substance is significantly lowered, ideally below $50$~mK.

A major disadvantage of ADR compared to $^{3}$He–$^{4}$He dilution refrigeration has been its incapability of continuous cooling, which is being addressed by the recent developments in continuous ADR~\cite{shirron581,bartlett582}. Commercial continuous refrigerators based on ADR are now available~\cite{kiutra}, suggesting that ADR has the potential to become a major refrigeration technology.

Even several decades after the first realization of magnetic refrigeration down to the mK range, hydrated paramagnetic salts with very low ordering temperatures are commonly used as cooling substances in ADR refrigerators. In the paramagnetic salts, the spins are very much diluted and are almost non-interacting, resulting in a low volumetric cooling power and low thermal conductivity. By contrast, systems with higher number of magnetic ions per unit volume may exhibit stronger exchange interactions, resulting in magnetic order. The magnetic ordering typically prevents the system from cooling down adiabatically to lower temperatures.

Even in almost-ideal paramagnets, there may be a very weak magnetic interaction $J$ through which the neighboring spins can interact weakly even at $H=0$, resulting in a small internal field. This can lead to a tiny Zeeman splitting ($\Delta_0$) and magnetic ordering at a temperature of the same energy scale. Magnetic ordering suddenly drops the entropy to zero and limits the final ADR temperature to $T_f \sim \Delta_0 \sim J $, since the entropy difference between $ H = 0 $ and a finite $H$ is an important parameter for ADR~\cite{tokiwa1}. A perfect paramagnet with zero interaction and maximum entropy at $H=0$ close to $ 0 $~K is an ideal refrigerant. But in real paramagnetic materials, the presence of a finite weak interaction is inevitable.

Recently the disordered quantum magnets came into focus. Since entropy accumulates near quantum critical points and/or due to magnetic frustration, new strategies for obtaining efficient mK-ADR were explored
~\cite{tokiwa1,liu1,kleinhans01752,wolf6862,jang1,brasiliano103002,tokiwa1600835,evangelisti6606,baniodeh1,zhitomirsky104421,wolf142112,hu125225}. 
In frustrated quantum magnets, the enhanced quantum fluctuations suppress the long-range ordering, despite the strong spin interactions, giving rise to a shift of entropy towards low temperatures, that is crucial for obtaining lower ADR end temperatures~\cite{tokiwa1}. For the efficiency of the cooling substance, the available entropy change per volume is an essential parameter~\cite{wikus150}. The size of the spins and their density constitute to the volumetric entropy density which should be as high as possible.

\begin{figure}
	\includegraphics[scale=0.26]{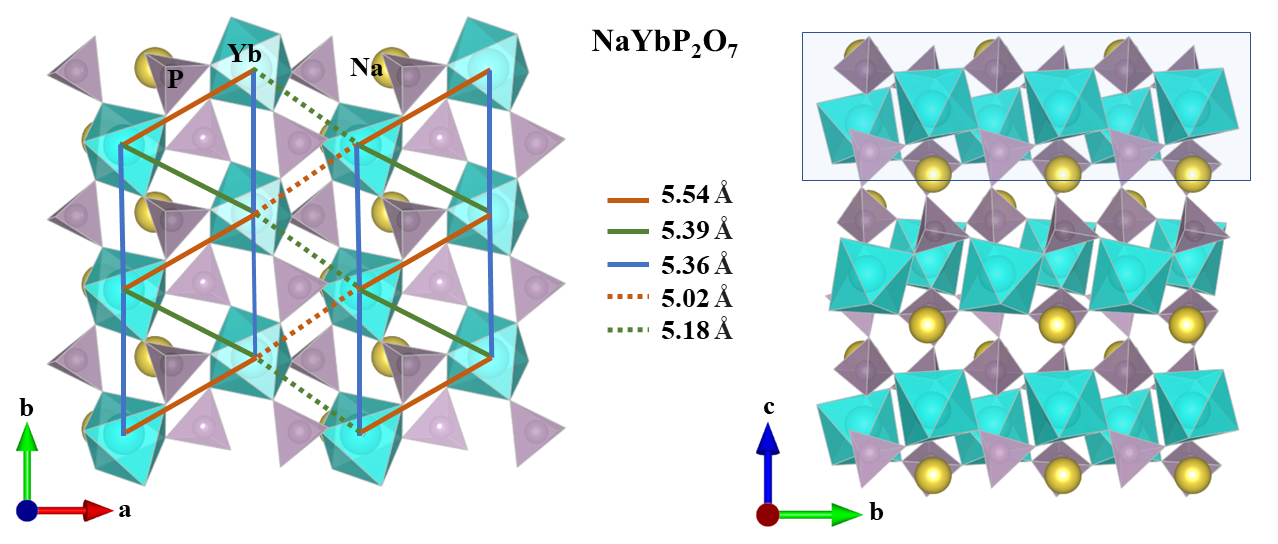}
	\caption{Crystal structure of NaYbP$_2$O$_7$ in the $ab$-plane and along the $c$-direction. The YbO$_6$ octahedra are linked via PO$_4$ tetrahedra and span a distorted triangular grid (left panel). One such triangular layer is highlighted by a rectangular box in the right panel. Different Yb-Yb distances are color-coded.}
	\label{Fig1}
\end{figure}

\begin{figure}
	\includegraphics[scale=0.26]{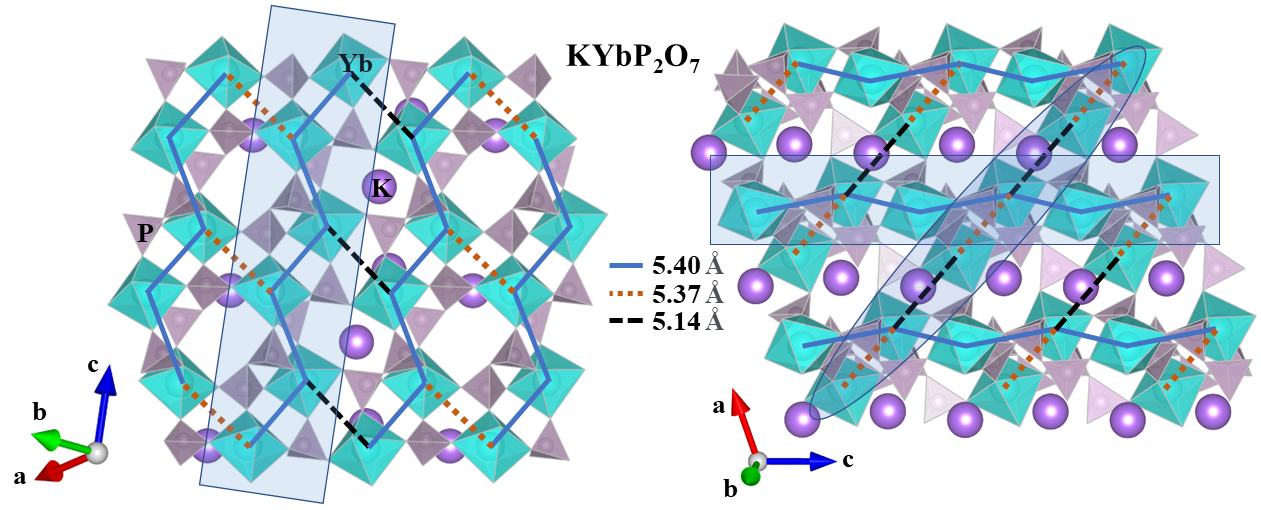}
	\caption{Crystal structure of KYbP$_2$O$_7$. The YbO$_6$ octahedra are linked via PO$_4$ tetrahedra forming uniform chains running along the $c$-axis (highlighted by a rectangular box) which are interconnected by alternating chains running in the $ac$-plane (highlighted by an ellipse in the right panel). This forms a three-dimensional distorted hyper-honeycomb-like network. Different Yb-Yb distances are color-coded.}
	\label{Fig2}
\end{figure}

Traditional mK-ADR substances such as CrK(SO$_4$)$_2~\cdot~12$H$_2$O (CPA)~\cite{daniels243} incorporate water of crystallization for separation of magnetic ions. Hence, the magnetic interactions are sufficiently small, resulting in very low magnetic ordering temperatures, like 30 mK for CPA. However, these salts have high water vapor pressure and, therefore, cannot be evacuated. This requires vacuum-tight encapsulation of the active material in a protective container. In order to ensure sufficient thermal coupling of very poorly thermally conductive salts, it is also necessary to grow them as single crystals within a very fine wire network~\cite{wikus150}. This complex production of encapsulated cooling units leads to high costs. Furthermore, they must not be heated due to the lack of chemical stability. This means that they cannot be thoroughly baked above $100 ^\circ$C for ultra-high-vacuum (UHV) applications.

Oxide quantum magnets, on the other hand, are stable upon heating and evacuation. Recently, it was discovered that potassium-barium-ytterbium borate KBaYb(BO$_3$)$_2$ can be used as an inert, easy-to-produce and inexpensive cooling substance for UHV-compatible adiabatic demagnetization cooling \cite{tokiwa1}. Magnetic order in KBaYb(BO$_3$)$_2$ is hindered by both geometrical frustration on the triangular lattice as well as the statistical mixing of K$^+$ and Ba$^{2+}$, resulting in uneven electric fields acting on the Yb$^{3+}$ ions, eventually causing a wide distribution of magnetic couplings. Isostructural KBaGd(BO$_3$)$_2$ shows magnetic order below $T_N=263$~mK in zero field, reaches a minimal ADR temperature of 122~mK and warm-up time of 8 hours in the PPMS setup~\cite{jesche23}. Dipolar and magnetic exchange couplings are of similar magnitude in this system~\cite{jesche23,xiang23}. Taking into consideration that dipolar and exchange interactions are proportional to the square of the magnetic moment and using the measured saturation moments for KBaYb(BO$_3$)$_2$ and KBaGd(BO$_3$)$_2$, suggests a magnetic order at 9~mK in the former~\cite{jesche23}.

In this work, we report mK magnetization, specific heat and ADR measurements on the geometrically frustrated quantum magnets NaYbP$_2$O$_7$ and KYbP$_2$O$_7$. Similar to 
KBaYb(BO$_{3}$)$_{2}$~\cite{tokiwa1}, the two materials are chemically stable and UHV-compatible. A direct comparison of the ADR performance with KBaYb(BO$_{3}$)$_{2}$ under similar conditions indicates for NaYbP$_2$O$_7$ a similar ADR end temperature but 30\% longer hold time and for KYbP$_2$O$_7$ a 20\% lower ADR end temperature at the cost of a 12.5\% shorter hold time. Thus, the two diphosphates are superior candidates for UHV-compatible mK-ADR.

\section{Methods}
\textbf{Synthesis:} Polycrystalline samples of $A$YbP$_{2}$O$_{7}$ ($A$= Na, K) are synthesized by the conventional solid-state reaction technique by annealing the stoichiometric mixture of Na$_2$CO$_3$/ K$_2$CO$_3$  (99.99\%), Yb$_2$O$_3$ (99.99\%), and NH$_{4}$H$_{2}$PO$_{4} $ (99.99\%) in an alumina boat at 650~$^\circ$C for NaYbP$_{2}$O$_{7}$ and 600~$^\circ$C for KYbP$_{2}$O$_{7}$ for a duration of $48$~h with one intermediate grinding and pelletization. The synthesis procedures for the KBaYb(BO$_{3}$)$_{2}$ pellet used for the comparison of ADR performances are described in Ref.~\onlinecite{tokiwa1}.

\textbf{Powder X-Ray diffraction:} Phase purities of the samples were confirmed by powder x-ray diffraction (XRD, PANalytical powder diffractometer with Cu$K_{\alpha}$ radiation, $\lambda_{\rm ave} = 1.54182$~\AA) at room temperature. Rietveld refinement of the observed XRD patterns was performed using the \verb|FullProf|~package~\cite{Carvajal55} (see Fig.~\ref{Fig3}), taking the initial parameters from Ref.~\onlinecite{horchani33,ferid385}.

\textbf{DC magnetization:} DC magnetization ($M$) was measured as a function of the temperature ($T$) down to 0.4~K and the applied magnetic field ($H$) up to 7~T in a MPMS-3 Quantum Design SQUID magnetometer with $^{3} $He option.

\textbf{Specific heat:}
Specific heat $C_{\rm p}(T)$ was measured using the heat capacity option of a Physical Property Measurement System (PPMS) manufactured by Quantum Design. For the low temperature ($0.4 $~K~$\leq T \leq 2.2$~K) $C_{\rm p}$ measurements the $^{3}$He option was used.

For strong thermal coupling, the specific heat measurements were performed on pellets made from sodium/potassium diphosphates (grain size $10-50~\mu $m) mixed with fine silver powder ($1~\mu $m) in a mass ratio of $1:1$. In order to extract the specific heat of the sample, we subtracted the Ag-contribution from the measured data. In a magnetic insulator, the specific heat $C_{\rm p}$ contains significant contributions from the phonon excitations ($C_{\rm ph}$) and  the magnetic lattice ($C_{\rm m}$). At high temperatures, $C_{\rm p}(T)$ is entirely dominated by $C_{\rm ph}$, while at low temperatures, it is dominated by $C_{\rm m}$. In order to estimate the phonon part of the specific heat, the $C_{\rm p}(T)$ data at high temperatures $(T > 25$~K) were described phenomenologically by a polynomial $C_{\rm ph}(T) = aT^3 + bT^5 + cT^7$,
where $a$, $b$, and $c$ are appropriate coefficients. Similar procedures have been used previously and have shown to be efficient for estimating $C_{\rm ph}$ in cases where heat capacity data of non-magnetic analogue compounds are not available~\cite{Guchhait104409,Nath054409,Matsumoto9993,Lancaster094421}. The fit was extrapolated down to low-temperatures, and the magnetic specific heat $C_{\rm m}$ was obtained by subtracting the obtained $C_{\rm ph}$ values from the experimental $C_{\rm p}$ data. Magnetic entropy $S_{\rm m}$ was estimated by integrating $C_{\rm m}(T)/T$ from $0.4$~K to high-temperatures as $S_{\rm m}(T) = \int_{0.4~K}^{T}\frac{C_{\rm m}(T')}{T'}dT'$.

\textbf{Adiabatic Demagnetization Refrigeration (ADR):} Using an ADR material in a commercial PPMS in order to achieve temperatures below $1.8$ K without using $^{3}$He presents an immediate practical application. For the cooling experiment, we used a $ \sim 3.5 $~g cylindrical pellet of $6$~mm thickness and $15$~mm diameter containing equal weights of $A$YbP$_{2} $O$_{7}$ (grain diameter $10–50~\mu$m) and silver powder (grain diameter $1~\mu$m). Silver powder is used to improve the thermal conductivity within the pellet as $A$YbP$_{2}$O$_{7}$ is insulating in nature. Also, we sintered the pressed pellet at $600~^\circ$C to further improve the thermal conductivity.

We performed the ADR experiments in the PPMS with a similar set-up as described in~\cite{tokiwa1}. A sample stage is constructed in which the pellet is mounted on a plastic straw. The sample is thermally isolated from the heat bath. An RuO$_{2}$ resistor is glued as a thermometer on the pellet and is connected using thin resistive manganin wires to minimize the heat flow. A metallic cap is used as a shield to minimize the effect of surrounding thermal radiation. PPMS is kept in the high-vacuum mode (pressure $< 10^{-4}$~mbar), and the bath temperature is set to $2$~K. Then a magnetic field of $5$~T is applied. Through the weak thermal link via manganin wires, the pellet slowly reaches $2$~K. After stabilizing at $2$~K, the magnetic field of $5$~T is swept to zero at a rate of $0.15$~T~min$^{-1}$. At zero field the pellet reaches the lowest temperature and then slowly up back to $2$~K by the slow heat flow from the bath. We also performed ADR experiments with a KBaYb(BO$_{3}$)$_{2}$ pellet from~\cite{tokiwa1} utilizing the same thermometer and parameters as in case of $A$YbP$_{2}$O$_{7}$. This allows a direct comparison of the ADR performances of two diphosphates with that of the previously reported borate.

\section{Results}
\subsection{Structure}

Both NaYbP$_{2}$O$_{7}$ and KYbP$_{2}$O$_{7}$ crystallize in a monoclinic lattice with the space group $P21/c$ (No. 14) but form different structure types. The crystal structure of NaYbP$_{2}$O$_{7}$ consists of distorted YbO$_{6}$ octahedra, which are corner-shared with PO$_{4}$ tetrahedra, forming a magnetic Yb triangular layer in the crystallographic $ab$-plane. These triangular planes are well separated by two non-magnetic PO$_{4}$ units and Na ions along the crystallographic $c$-direction (see Fig.~\ref{Fig1}). In KYbP$_{2}$O$_{7}$, on the other hand, the YbO$_{6}$ octahedra are connected by PO$_{4}$ tetrahedra forming uniform chains running along the $c$-axis and are connected by alternating chains in the $ac$-plane. It forms a three-dimensional distorted hyper-honeycomb-like network (see Fig.~\ref{Fig2}).

\subsection{Powder x-Ray diffraction}
\begin{figure}
	\includegraphics[scale=0.32]{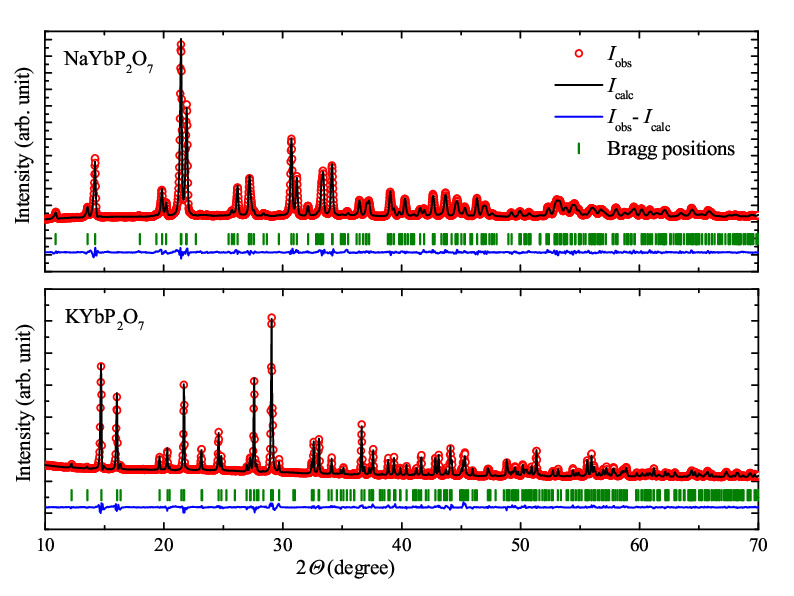}
	\caption{Powder x-ray diffraction pattern (open red circles) for NaYbP$_2$O$_7$ (upper panel) and KYbP$_2$O$_7$ (lower panel) at room temperature. The solid line represents the Rietveld refinement, with the vertical bars showing the expected Bragg peak positions and the lower solid blue line representing the difference between observed and calculated intensities.}
	\label{Fig3}
\end{figure}
Figure~\ref{Fig3} shows the powder XRD pattern of NaYbP$_{2}$O$_{7}$ and KYbP$_{2}$O$_{7}$ at room temperature, along with the Rietveld refinement. For NaYbP$_{2}$O$_{7}$, the refinement uses the monoclinic space group $P2_1/c$ (No. 14) $[$settings $P2_1/n$ (unique axis-$b$)$]$, taking the initial parameters from Ref.~\onlinecite{ferid385}. The goodness of fit is $\chi^2 = 4.72$. The obtained lattice parameters are $a=9.0215(1)$~\AA, $b=5.3599(1)$~\AA, $c=12.7816(1)$~\AA, $\beta=103.1655(1)^\circ$, and $V_{\rm cell}\simeq 601.81(1)$~\AA$^3$. These values are in close agreement with reported values~\cite{ferid385}.
For KYbP$_{2}$O$_{7}$, the refinement is performed using the monoclinic space group $P2_1/c$ (No. 14) $[$settings $P2_1/c$ (unique axis-$b$)$]$, taking the initial parameters from Ref.~\onlinecite{horchani33}. The goodness of fit is $\chi^2 = 8.78$. The obtained lattice parameters are $a = 7.5500(1)$~\AA, $b = 10.8306(1)$~\AA, $c = 8.5492(1)$~\AA, $\beta = 106.7200(1)^\circ$, and $V_{\rm cell}\simeq 669.52(1)$~\AA$^3$. These values are in close agreement with reported values~\cite{horchani33}.
The unit cell volume of NaYbP$_{2}$O$_{7}$ is small compared to KYbP$ _{2}$O$_{7}$ because the ionic radius of Na$^+$ (1.18~\AA) is smaller than that of K$^+$ (1.51~\AA)~\cite{Shannon}.

\subsection{Magnetization}
\begin{figure}
	\includegraphics[scale=0.32]{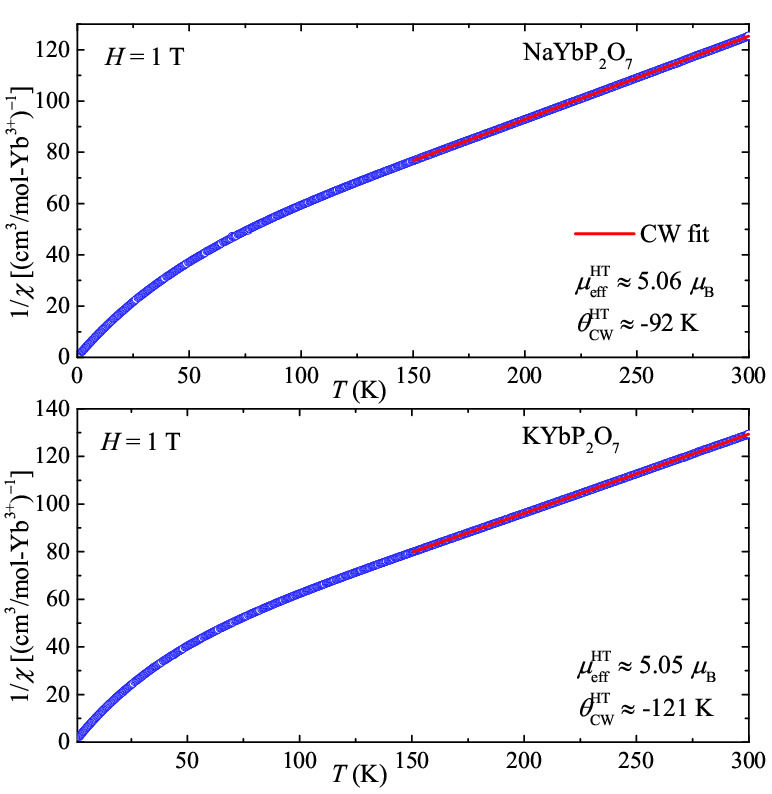}
	\caption{Inverse magnetic susceptibility ($1/\chi$) data measured at $1$~T as a function of temperature ($T$) for NaYbP$_{2}$O$_{7}$ (upper panel) and KYbP$_{2}$O$_{7}$ (lower panel). The solid line represents the high-temperature CW-fit according to Eq.~\ref{cw}.}
	\label{Fig4}
\end{figure}
\begin{figure*}
	\includegraphics[scale=0.33]{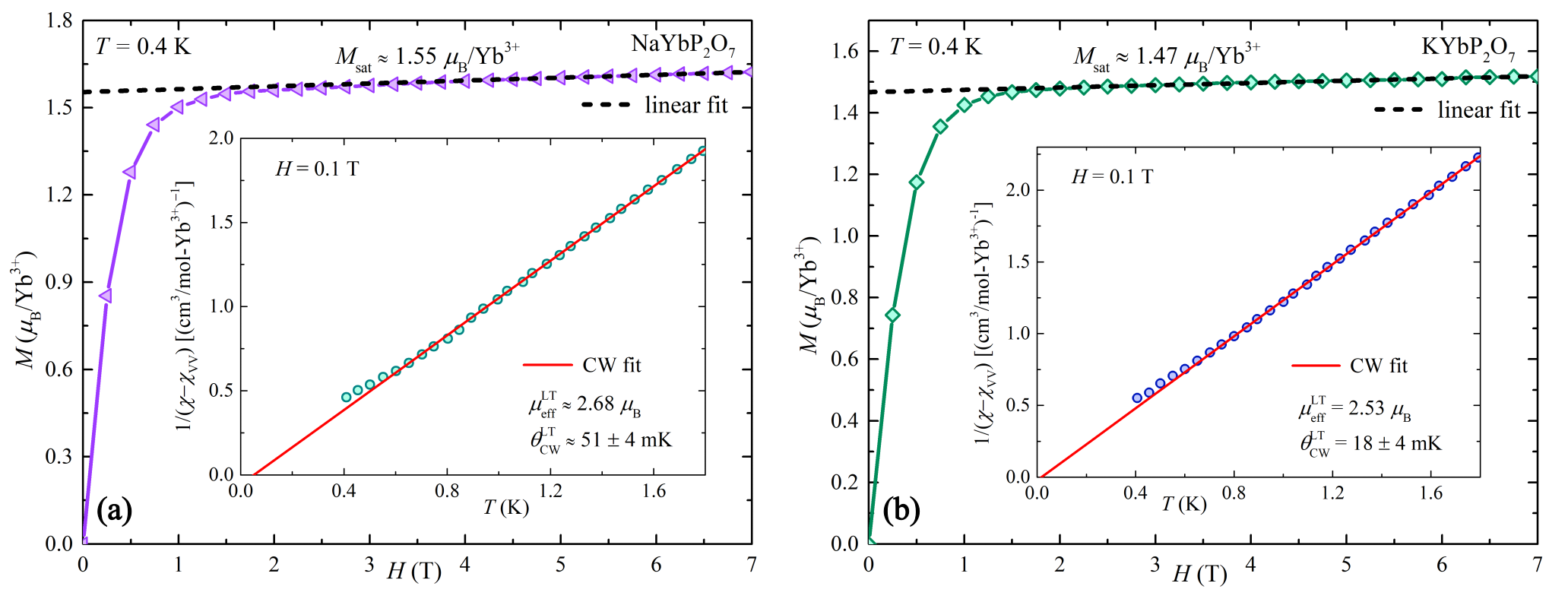}
	\caption{Isothermal magnetization $M(H)$ measured at $0.4$~K for NaYbP$_2$O$_7$ (panel (a)) and KYbP$_2$O$_7$ (panel (b)). Dashed lines represent linear contributions (see text). The insets show the low-temperature inverse magnetic susceptibility after subtracting $\chi_{\rm VV}$ along with the CW-fit for NaYbP$_2$O$_7$ [inset of (a)] and KYbP$_2$O$_7$ [inset of (b)].}
	\label{Fig5}
\end{figure*}

The temperature-dependent magnetic susceptibility $\chi(T)=M/H$ data of NaYbP$_{2}$O$_{7}$ and KYbP$_{2}$O$_{7}$ were measured at applied fields $H=0.1$~T and $1$~T. No evidence of any long-range magnetic order was observed down to $0.4 $~K. At high temperatures (above $150$~K), $1/\chi(T)$ can be well fitted with the modified Curie-Weiss (CW) expression (see Fig.~\ref{Fig4})
\begin{equation}\label{cw}
	\chi(T) = \chi_0 + \frac{C}{T - \theta_{\rm CW}},
\end{equation}
where $\chi_0$ is the temperature-independent contribution consisting of the diamagnetic susceptibility of core electron shells ($\chi_{\rm core}$) and the van-Vleck paramagnetic susceptibility ($\chi_{\rm vV}$) of the open shells of the Yb$^{3+}$ ions. The second term in Eq.~\ref{cw} is the CW law with the CW temperature $\theta_{\rm CW}$ and Curie constant $C=N_{\rm A} \mu_{\rm eff}^2/3k_{\rm B}$. Here, $N_{\rm A}$ is Avogadro's number, $\mu_{\rm eff} = g\sqrt{J(J+1)}$$\mu_{\rm B}$ is the effective magnetic moment, $g$ is the Land\'{e} $g$-factor, $\mu_{\rm B}$ is the Bohr magneton, and $ J $ is the effective spin.

The fitting yields the parameters $\chi_{0}^{\rm HT} \simeq -1.73 \times 10^{-4}$ cm$^{3} $/mol, $ C_{\rm HT} \simeq 3.20$~cm$^{3} $K/mol, and $\theta_{\rm CW}^{\rm HT} \simeq -92$~K for NaYbP$_{2}$O$_{7}$ and $\chi_{0}^{\rm HT} \simeq -2.48 \times 10^{-4}$cm$^{3}$/mol, $C_{\rm HT} \simeq 3.18$~cm$^{3}$K/mol, and $\theta_{\rm CW}^{\rm HT} \simeq -121$~K for
KYbP$ _{2} $O$ _{7} $. The resulting effective moment $\mu_{\rm eff}^{\rm HT}$ was calculated to be $ \sim 5.06~\mu_{\rm B}$ and $ \sim 5.05~\mu_{\rm B}$ for NaYbP$_{2}$O$_{7}$ and KYbP$_{2}$O$_{7}$, respectively.  These values of $\mu_{\rm eff}^{\rm HT}$ are in good agreement with the expected value of $4.54~\mu_{\rm B}$ for Yb$ ^{3+}$ ($J = 7/2$, $g = 8/7$) ion in the $4f ^{13}$ configuration.

A change in slope was observed in $1/\chi(T)$ data below $50$~K, arising from thermal depopulation of excited crystal electric field (CEF) levels. Similar behavior was observed in several other Yb$^{3+}$-based spin systems. Typically, the combination of the spin orbit coupling and the CEF leads to a Kramers doublet ground state for the Yb$^{3+}$ ion, and the low temperature properties can be described by an $S_{\rm eff}=1/2$ ground state ~\cite{Paddison117,Bordelon1058,Ranjith224417,Baenitz220409,Schmidt214445,Ranjith180401}.

The magnetization isotherms $M(H)$ of KYbP$_{2}$O$_{7}$ (see Fig.~\ref{Fig5}a) and NaYbP$_{2}$O$_{7}$ (see Fig.~\ref{Fig5}b) measured up to 7~T at $T=0.4$~K show saturation around 1.5~T. The saturation values $M_{\rm sat}$ are estimated by fitting the high-field regions above 5~T with straight lines reflecting $\chi_0 H$ and extrapolating the lines back to zero field. This yields $\chi_0 \simeq 5.51\times 10^{-3}$~cm$ ^{3} $/mol and $M_{\rm sat} \simeq 1.55~\mu_{\rm B}$ for NaYbP$_{2}$O$_{7}$ and $\chi_0 \simeq 4.11\times 10^{-3}$ cm$^{3}$/mol and $ M_{\rm sat} \simeq 1.47~\mu_{\rm B}$ for KYbP$_{2}$O$_{7}$. Using $\mu_{\rm sat}=g_{\rm eff}S_{\rm eff}~\mu_{\rm B}$ gives
$g_{\rm eff}$ values of $\simeq 3.10 $ for NaYbP$_{2}$O$_{7}$ and $\simeq2.94 $ for KYbP$_{2}$O$_{7}$.

After subtracting $\chi_0$, the $\chi(T)$ data below $2$~K are fitted by the CW law $\chi(T)=C/(T-\theta_{\rm CW})$. The fitting yields $ \theta^{\rm LT}_{\rm CW} \simeq (51\pm 4)$~mK and $\mu_{\rm eff}^{\rm LT} \simeq 2.68~\mu_{\rm B}$ for NaYbP$ _{2} $O$ _{7} $ and $ \theta^{\rm LT}_{\rm CW} \simeq (18\pm 4) $~mK and $\mu_{\rm eff}^{\rm LT} \simeq 2.53~\mu_{\rm B}$ for KYbP$ _{2} $O$ _{7} $. The obtained effective moments are in good agreement with $\mu_{\rm eff} = g_{\rm eff}\sqrt{S_{\rm eff}(S_{\rm eff}+1)}~\mu_{\rm B}$ for a pseudo spin-$ 1/2 $ ground state with $g_{\rm eff}$ values of $ \sim 3.10 $ for NaYbP$ _{2} $O$ _{7} $ and $ \sim 2.92 $ for KYbP$ _{2} $O$ _{7} $, consistent with those estimated from $ M_{\rm sat} $.

Very small values of $\theta^{\rm LT}_{\rm CW}$ at low-$T$ suggest the lack of significant exchange interactions between Yb$^{3+}$ magnetic moments. According to mean field theory, $\theta_{\rm CW}$ is the sum of all possible exchange interactions~\cite{domb1964}. Since there are many possible interaction pathways in the structure, as shown by Fig.~\ref{Fig1} and Fig.~\ref{Fig2}, it is also possible that the competing ferromagnetic and antiferromagnetic exchange interactions cancel each other, resulting in very small values of $\theta^{\rm LT}_{\rm CW}$.

\subsection{Specific heat}
\begin{figure*}
	\includegraphics[scale=0.33]{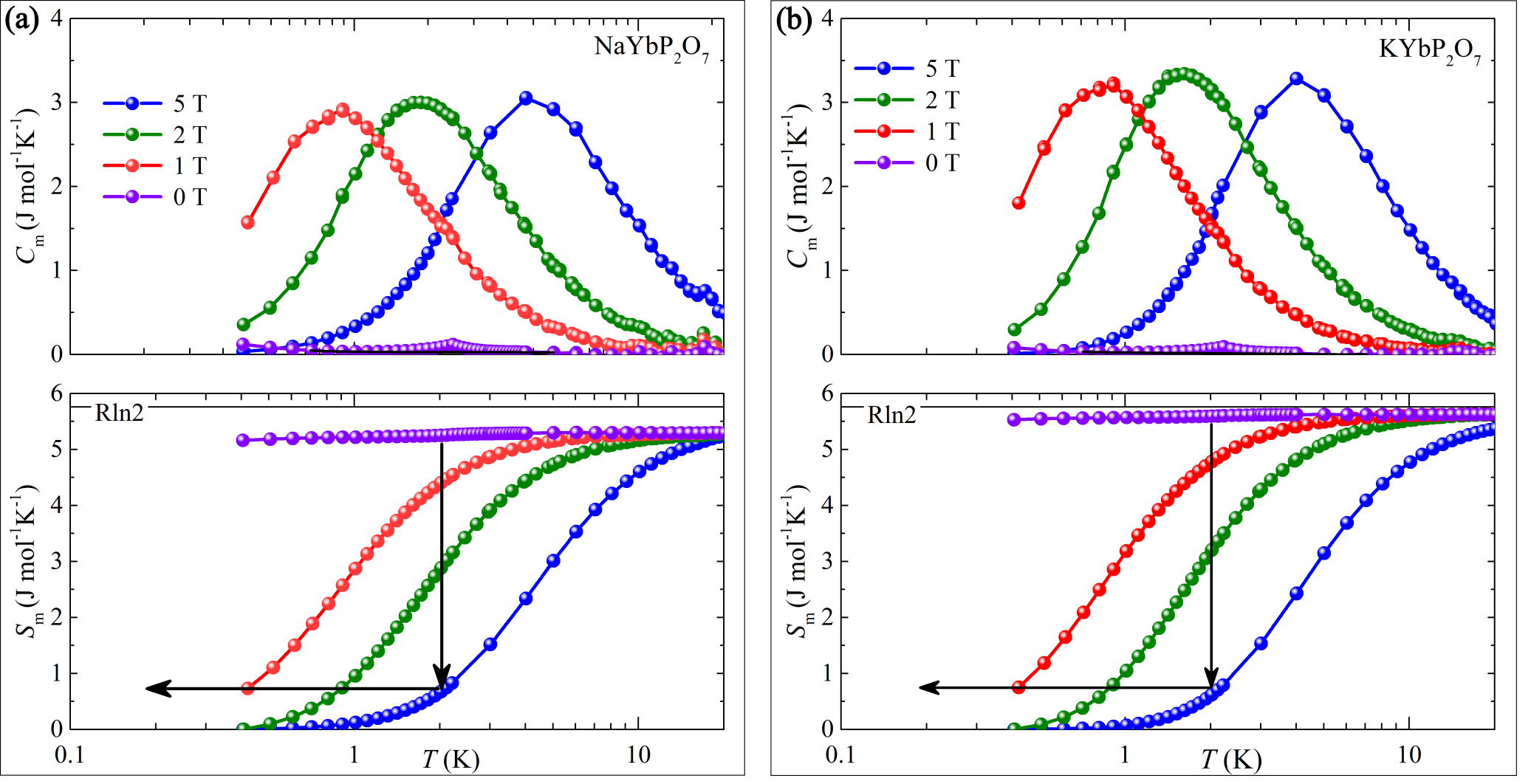}
	\caption{Low-temperature magnetic specific heat $(C_{\rm m})$ of NaYbP$_2$O$_7$ (upper panel of (a)) and KYbP$_2$O$_7$ (upper panel of (b)) at several external magnetic fields. Phonon contribution was subtracted from the raw data using a polynomial fit. Magnetic entropy $(S_{\rm m})$ of NaYbP$_2$O$_7$ (lower panel of (a)) and KYbP$_2$O$_7$ (lower panel of (b)) calculated by integrating $C_{\rm m}/T $. At $ 0 $~T and $ 1 $~T fields, the entropy is vertically shifted to $ 2 $~T and $ 5 $~T values to match $R\ln 2$ expected for the Kramers doublet. Two arrows show the ADR process.}
	\label{Fig6}
\end{figure*}

The specific heat analyses of NaYbP$_{2}$O$_{7}$ and KYbP$_{2}$O$_{7}$ at different applied fields are shown in various panels of Fig.~\ref{Fig6}, suggesting their ADR potentials.
The specific heat data of NaYbP$_{2}$O$_{7}$ and KYbP$_{2}$O$_{7}$ show no signatures of magnetic long-range ordering down to $ 0.4 $~K. In  zero-field, the specific heat shows an increase towards lower temperatures, suggesting the entropy accumulation associated with the Kramers doublet of Yb$^{3+}$. This doublet is split by the magnetic field into $ j_{z} =+1/2$ and $ j_{z} =-1/2$ levels, causing a Schottky anomaly that shifts toward higher temperatures with increasing field. 

For NaYbP$_{2}$O$_{7}$ and KYbP$_{2}$O$_{7}$, the resulting magnetic entropy ($S_m$) saturates about $5.3$~J/mol~K and $5.6$~J/mol~K, respectively (see Fig.~\ref{Fig6}). The values agree quite well with the expected theoretical value $S_{\rm m} = R \ln( 2J+1)$ of 5.76~J/mol~K for the Kramers doublet with $ J_{\rm eff} =1/2 $. At $ 0 $~T and $ 1 $~T, the entire entropy of $R\ln 2 $ associated with the lowest Kramers doublet of Yb$ ^{3+} $ could not be recovered as we did not measure below $ 0.4 $~K. Higher fields shift the entropy toward higher temperatures, so that at $ 2 $~T, almost complete entropy of $R\ln 2$ can be recovered above $ 0.4 $~K. The entropy data at $ 0 $~T and $ 1 $~T, are vertically shifted to match values at higher-fields ($ 2 $~T and $ 5 $~T). At $ 5 $~T, only $\sim10$\% of $R\ln 2$ remains at $T=2$~K. Therefore, by adiabatic demagnetization starting from $ 2 $~K and $ 5 $~T, very low temperatures can be attained as indicated by arrows in Fig.~\ref{Fig6}.

In the zero-field specific heat data of NaYbP$ _{2} $O$ _{7} $ and KYbP$ _{2} $O$ _{7} $, a tiny anomaly at $ \sim2.23 $~K is found, arising from a Yb$_2 $O$_3 $ impurity contribution~\cite{moon1383}. Integrating the entropy related to this impurity peak yields a negligible fraction of  $\sim 0.6\%$ of $R \ln 2$ related to the impurity contribution for both compounds. 


\subsection{Adiabatic Demagnetization Refrigeration (ADR)}
\begin{figure}
	\includegraphics[scale=0.32]{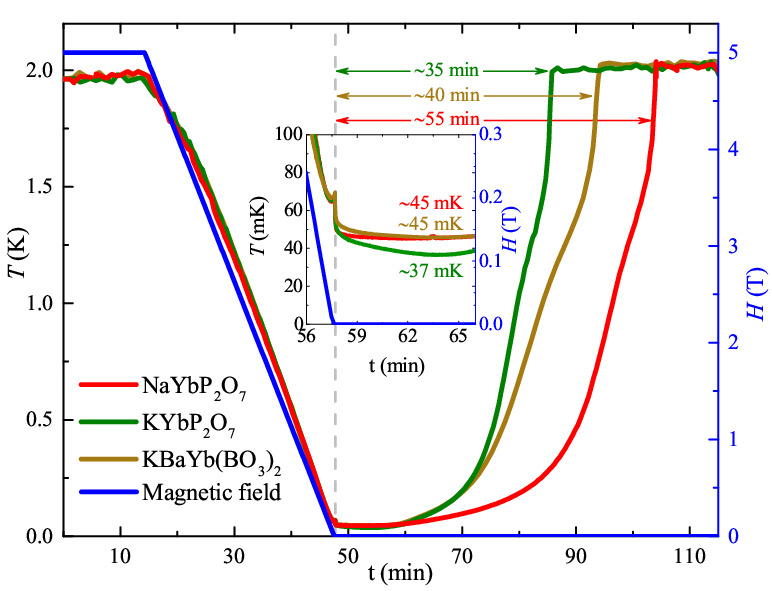}
	\caption{ADR performance comparison of NaYbP$_{2}$O$_{7}$ (red), KYbP$_{2}$O$_{7}$ (green), and KBaYb(BO$_{3}$)$_{2}$ (brown) in the commercial physical property measurement system (PPMS). The sample is slowly cooled to $ T = 2 $~K at $ H = 5$~T through the weak thermal link to the PPMS puck kept at $ 2 $~K and the 'high vacuum mode' of the PPMS was employed in order to achieve thermal decoupling. Subsequently, magnetic field is swept from $5$~T to zero at a rate of $0.15$~T min$ ^{-1}$ and the temperature change with time t is recorded. The blue curve shows the respective magnetic field. 
}
	\label{Fig7}
\end{figure}
The comparison of the cooling performances of NaYbP$ _{2} $O$ _{7} $ (red) and KYbP$ _{2} $O$ _{7} $ (green) with KBaYb(BO$ _{3} $)$ _{2} $ (yellow) in the commercial PPMS are shown in Fig~\ref{Fig7}. The sample temperatures are plotted on the left $ y $-axis, while the external magnetic field is plotted on the right $ y $-axis. During demagnetization, the lowest temperature attained by KYbP$ _{2} $O$ _{7} $ is $ \sim 37 $~mK while the lowest temperatures attained by NaYbP$ _{2} $O$ _{7} $ and KBaYb(BO$ _{3} $)$ _{2} $ are both $ \sim 45 $~mK. A spike in the sample temperature around $ 70 $~mK is observed, possibly due to flux pinning of the superconducting magnet.

The time required for NaYbP$ _{2} $O$ _{7} $ to relax back from the lowest temperature to $ 2 $~K is $\sim 55$~min, whereas the warming time for KYbP$ _{2} $O$ _{7} $ is shorter, i.e., about $\sim 35$~min. The warming time of NaYbP$ _{2} $O$ _{7} $ is much longer than that of KBaYb(BO$ _{3} $)$ _{2} $ ($\sim 40 $~min). Note that the hold-times at low temperatures could be significantly enhanced by using a thermal shield and precooling ADR station for all connected wires. But already in this simple setup, the warm-up times are sufficient to use the setup for instance to perform temperature dependent electrical resistance measurements on small samples. Importantly, the obtained base temperatures are much lower as for the commercially available ADR option for the PPMS, which uses CPA and reaches a minimum temperature of $100 $~mK~\cite{QDADR}.

\section{Discussion}

\begin{table}[h]
	\caption{Comparison of important parameters of different mK ADR materials: $T_{\rm m}$ is the magnetic ordering temperature, $S_{\rm GS}$ is the entropy of the ground-state multiplet, and $R$ is the universal gas constant. The abbreviations are MAS = Mn(NH$_{4}$)$_{2}$(SO$_{4}$)$_{2}\cdot6$H$_{2}$O (manganese ammonium sulfate), FAA = NH$_{4}$Fe(SO$_{4}$)$ \cdot $$12$H$_{2}$O (ferric ammonium alum), CPA = KCr(SO$_4$)$ \cdot 12$H$_{2}$O (chromium potassium alum), CMN = Mg$_{3}$Ce$_{2}$(NO$_{3}$)$_{12}\cdot24$H$_{2}$O (cerium magnesium nitrate).}
	\label{ADR_parameters}
	\begin{tabular}{ccccc}
		\hline \hline
		&&ADR materials&&\\
		\hline
		Material & $ T_{\rm m} $ &mag. ion/vol.& $ S_{ \rm GS}$ & $ S_{\rm GS} $/vol. \\
		& (mK) & (nm$ ^{-3} $) &  & [mJ/(K cm$ ^{3} $)] \\
		\hline
		MAS~\cite{Vilches509} & $ 170 $ & $ 2.8 $ & $R\ln 6$ & $ 70 $ \\
		FAA~\cite{Vilches509} & $ 30 $ & $ 2.1 $ & $R\ln 6$ & $ 53 $\\
		CPA~\cite{daniels243} & $ 10 $ & $ 2.2 $ & $R\ln 4$ & $ 42 $\\
		CMN~\cite{Fisher5584} & $ 2 $ & $ 1.7 $ & $R\ln 2$ & $ 16 $\\
		YbPt$ _{2} $Sn$ _{12} $~\cite{jang1} & $ 250 $ & $ 12.9 $ & $R\ln 2$ & $ 124 $\\
		Yb$ _{3} $Ga$ _{5} $O$ _{12} $~\cite{brasiliano103002} & $ 54 $ & $ 13.2 $ & $R\ln 2$ & $ 124 $\\
		KBaYb(BO$_{3}$)$_{2}$~\cite{tokiwa1} & $ <22 $& $ 6.7 $& $R\ln 2$ & $ 64 $\\
		NaYbP$ _{2}$O$_{7}$ & $ <45 $ & $ 6.6 $ & $R\ln 2$ & $ 64 $\\
		KYbP$ _{2}$O$_{7}$ & $ <37 $ & $ 6.0 $ & $R\ln 2$ & $ 57 $\\
		\hline \hline
	\end{tabular}
\end{table}

The low-temperature properties of both NaYbP$ _{2}$O$_{7}$ and KYbP$ _{2}$O$_{7}$ can be well described by an effective spin-$ 1/2 $ ground state expected for Yb$ ^{3+} $-based quantum magnets. Very small CW temperatures obtained from the analysis of the susceptibility data at low temperatures indicate the absence of significant interactions between the Yb-moments, which is advantageous to achieve low final temperatures in ADR measurements.

The compounds NaYbP$ _{2}$O$_{7}$ and KYbP$ _{2}$O$_{7}$ have proven to be very efficient for ADR applications because they do not contain water molecules and are stable to heating and evacuation. It can be seen from Fig.~\ref{Fig6} that almost all of the entropy ($R\ln 2 $) of the lowest Kramers doublet can be used for cooling, even at a moderate magnetic field of $ 5 $~T, similar to the conventionally used paramagnetic salts~\cite{pobell2007}.
The crucial parameters determining the efficiency of various known ADR materials are compared in Table~\ref{ADR_parameters}. The full entropy of the ground state is described as $ S_{\rm GS}= R\ln (2J+1) $, and the entropy density $ S_{\rm GS} $/vol. is calculated by dividing $ S_{\rm GS}$ by the unit cell volume. A large value of $ S_{\rm GS}$ is always beneficial because the magnetic entropy changes ($ \Delta S_{\rm m} $) of magnetic refrigerants act as the driving force of ADR. However, it is worth noting that for practical purposes, the relevant quantity is not the molar entropy but rather the volumetric entropy density $S_{\rm GS} $ of the material.

As can be seen from the Table.~\ref{ADR_parameters}, materials with high entropy density such as YbPt$_{2}$Sn or Yb$_{3}$Ga$_{5}$O$_{12}$ exhibit magnetic orderings at $ 250 $ mK and $ 54 $~mK, respectively, which limit their lowest attainable temperatures as the entropies drop to zero below the ordering temperatures. The high-temperature paramagnetic salts such as MAS and FAA, which have a larger magnetic entropy $R\ln 6$ are also affected by their much higher magnetic ordering temperatures. On the other hand, low transition temperature materials such as CPA and CMN have low magnetic moment density and hence low entropy density. A high entropy density usually contradicts a low magnetic ordering temperature.

In this context, it has been reported that the disordered quantum magnet KBaYb(BO$ _{3} $)$ _{2} $ exhibits these two mutually exclusive criteria, such as a high volumetric entropy density ($ 64 $~mJ/(K cm$ ^{3} $)) combined with a very low ordering at approximately 9~mK~\cite{jesche23}, allowing ADR to well below 20~mK~\cite{tokiwa1}. In this compound, the magnetic frustration and the structural randomness help to suppress the magnetic order. Because of its suitable properties, KBaYb(BO$_{3}$)$_{2}$ has been proven to be an excellent anhydrous ADR refrigerant. By demagnetizing $H=5$~T at $2$~K in PPMS, a minimal temperature of $\sim 40 $~mK was attained with pellets mixed of KBaYb(BO$ _{3} $)$ _{2} $ powder with Ag powder~\cite{tokiwa1}. Utilizing a better adiabatic setup in the dilution refrigerator with feedback control of the bath temperature following the sample temperature, KBaYb(BO$ _{3} $)$ _{2} $ cooled upon demagnetization starting at 5~T from 2~K to well below 20~mK~\cite{tokiwa1}.

NaYbP$ _{2} $O$ _{7} $ and KYbP$ _{2} $O$ _{7} $ have also high magnetic ion densities ($ \sim 6.6 $ and $ \sim 6 $~nm$ ^{-3} $) and volumetric entropy densities ($ \sim 64 $ and $ \sim 57 $~mJ/(K cm$ ^{3} $)). These values are comparable to those of KBaYb(BO$ _{3} $)$ _{2} $ but are much higher than those of the paramagnetic ADR salts for the mK application.
The comparative study of the ADR performance of AYbP$ _{2} $O$ _{7} $ and KBaYb(BO$ _{3} $)$ _{2} $ pellets under exactly similar conditions revealed a significantly lower minimal temperature of $ \sim37 $~mK for KYbP$ _{2} $O$ _{7} $, whereas NaYbP$ _{2} $O$ _{7} $ reaches the same minimum temperature of $ \sim45 $~mK as KBaYb(BO$ _{3} $)$ _{2} $ but with the advantage of a much longer warm-up time. Comparison with our previous work on KBaYb(BO$ _{3} $)$ _{2} $~\cite{tokiwa1} suggests that KYbP$ _{2} $O$ _{7} $ and NaYbP$ _{2} $O$ _{7} $ can also be cooled by demagnetization to significantly lower temperatures of at least 20~mK in better adiabatic conditions (as realized previously for the former utilizing a $ ^{3} $He-$ ^{4} $He dilution refrigerator).

Compared to commercially used conventional mK ADR coolants based on paramagnetic salts, $ A $YbP$ _{2} $O$ _{7} $ and KBaYb(BO$ _{3} $)$ _{2} $ are anhydrous compounds and hence stable at high-vacuum and high temperatures up to at least $ 600~^\circ $C. Therefore, an encapsulated installation is not required, making them user-friendly and suitable for UHV applications. Finally, excellent thermal contact can be achieved in easily prepared pellets by mixing powder samples with silver powder in $ 1:1 $ ratio. Overall, it has been shown that all three cooling substances are very well suited for UHV-compatible ADR down to $ 50 $~mK.

\section{Conclusion}

In summary, we have performed a comprehensive study on the low-temperature properties of two Yb-based quantum magnets NaYbP$ _{2} $O$ _{7} $ and KYbP$ _{2} $O$ _{7} $ and compared their mK-ADR performance with KBaYb(BO$_{3}$)$_{2}$. The low-temperature properties are well described by very weakly interacting $J_{\rm eff}=1/2$ Kramers doublets of the Yb$^{3+}$ ions. ADR experiments in the PPMS under comparable conditions for all three compounds confirm that all the three are highly suitable to achieve temperatures below 50~mK. With respect to KBaYb(BO$_{3}$)$_{2}$, KYbP$_{2}$O$_{7}$ yields a $20\%$ lower temperature (but $12.5\%$ shorter hold time), while for NaYbP$_{2}$O$_{7}$ a similar end temperature is combined with $30\%$ longer hold time compared to KBaYb(BO$_{3}$)$_{2}$. Both diphosphates are thus excellent new UHV-compatible mK ADR materials.


\acknowledgments
UA would like to acknowledge DST, India, for financial support bearing sanction (DST/INSPIRE/04/2019/001664). Work supported by the German Science Foundation through projects 107745057 (TRR80) and 514162746 (GE 1640/11-1). DDS thanks SERB, DST, and CSIR, Government of India, for financial support. We note that a German patent for the usage of $AB$P$_2$O$_7$ ($A$=alkaline metal, $B$=rare earth) for UHV compatible ADR to very low temperatures has been filed by the University of Augsburg (file reference DE 10 2023 106 074.0, March 10, 2023).

\bibliography{ref_NKYP}

\end{document}